\documentclass[a4paper,11pt]{article}
\pdfoutput=1

\usepackage{jinstpub}
\usepackage{caption}
\usepackage{textcomp}
\usepackage{multirow}
\usepackage{comment}

\title{Measurement of high-pressure xenon gas absorption in acrylic}

\author[a]{Heng Lin,}
\author[b]{Kaixiang Ni,}
\author[b]{Haiyan Du,}
\author[b,1]{Ke Han,\note{Corresponding authors.}}
\author[c]{Xiangdong Ji,}
\author[d]{Tao Li,}
\author[e]{Lingyin Luo,}
\author[b,f]{Shaobo Wang,}
\author[d]{Xiyv Yan,}
\author[b]{Xinning Zeng,}
\author[b]{Wenming Zhang}

\affiliation[a]{Tsung-Dao Lee Institute, Shanghai Jiao Tong University, Shanghai, 200240, China}
\affiliation[b]{School of Physics and Astronomy, Shanghai Jiao Tong University, MOE Key Laboratory for Particle Astrophysics and Cosmology,
Shanghai Key Laboratory for Particle Physics and Cosmology, Shanghai 200240, China}
\affiliation[c]{Department of Physics, University of Maryland, College Park, Maryland 20742, USA}
\affiliation[d]{School of Physics, Sun Yat-Sen University, Guangzhou 510275, China}
\affiliation[e]{School of Physics, Peking University, Beijing 100871, China}
\affiliation[f]{SJTU Paris Elite Institute of Technology, Shanghai Jiao Tong University, Shanghai 200240, China}

\emailAdd{ke.han@sjtu.edu.cn;}

\abstract{Acrylic is a popular structural material in experiments requiring low background because of its radio-purity, machinability, and mechanical strength.
However, its porosity may cause significant gas absorption and influence the detector stability in the long term.
The interaction between acrylic and other detector materials becomes one of the key concerns in the detector design.
In this paper, we carry out an experiment to measure quantitatively the absorption process of high-pressure xenon gas into acrylic.
A specific setup is designed for the measurement, and systematic measurements are done to obtain a result of the absorption amount: 0.98~g xenon into 332~g of acrylic.
}

\keywords{}

\arxivnumber{} 

\begin{document}
\maketitle
\flushbottom


\section{Introduction}
\label{sec:Introduction}

For particle experiments searching for rare events, radio-purity is one of the key concerns which directly constrains the sensitivity.
Extensive experimental efforts have been made to search, select, and clean materials with low radioactivity.
Poly(methyl methacrylate), also known as acrylic or PMMA, has been and is widely used in experiments requiring low background (SNO~\cite{SNO:1999crp}, DEAP~\cite{DEAP-3600:2017ker}, PandaX-III~\cite{Chen:2016qcd}, JUNO~\cite{JUNO:2015sjr}, DarkSide~\cite{DarkSide-20k:2017zyg}, etc.).
It can serve as the structural or light guiding material, because of its mechanical strength, superior radio-purity~\cite{JUNO:2021kxb,Nantais:2013}, transparency~\cite{Yang:2021ecv} as well as its chemical compatibility with other detector materials.

Despite its chemical stability, acrylic is physically very active.
Acrylic tends to trap water and gas easily due to its micro-structural porosity.
A water absorption ratio up to 0.4\% by weight has been reported~\cite{matbase}.
Consequently, profuse outgassing under vacuum for an extended time period~\cite{NEXT:2018wtg} may prolong the pumping process significantly and/or even introduce contaminants to detectors.
No specific study has been made on the physical interaction between acrylic and the noble gases which are often used in rare event search experiments.
For example, xenon is widely used in neutrinoless double beta decay experiments~\cite{Chen:2016qcd,KamLAND-Zen:2016pfg,EXO-200:2012pdt,NEXT:2012zwy} and dark matter direct detection experiments~\cite{PandaX-4T:2021bab, XENON:2017lvq, LZ:2015kxe}.
In this paper, we described the experimental design of a dedicated system to quantitatively measure the xenon absorption rate in acrylic under high pressure.

If an acrylic sample is sealed in a chamber filled with xenon gas, acrylic is expected to absorb the gas gradually.
The absorption process can be characterized by the pressure drop of xenon gas in the chamber.
A precise measurement of pressure drop over a time scale of days or even months requires careful temperature stability control and/or monitoring.
Multiple barometers together with temperature sensors offer precise measurements of the temperature-corrected pressure of major system segments.
Besides identifying the absorption amount from the pressure, the acrylic sample can also be weighed before and after the absorption to provide the cross-check result.
We describe our setup and measurement results of the pressure drop and weight methods in the next two sections respectively.

\section{Absorption measurement by pressure drop}
\label{sec:pressure}

\begin{figure}[tb]
  \begin{center}
    \includegraphics[width=1\linewidth]{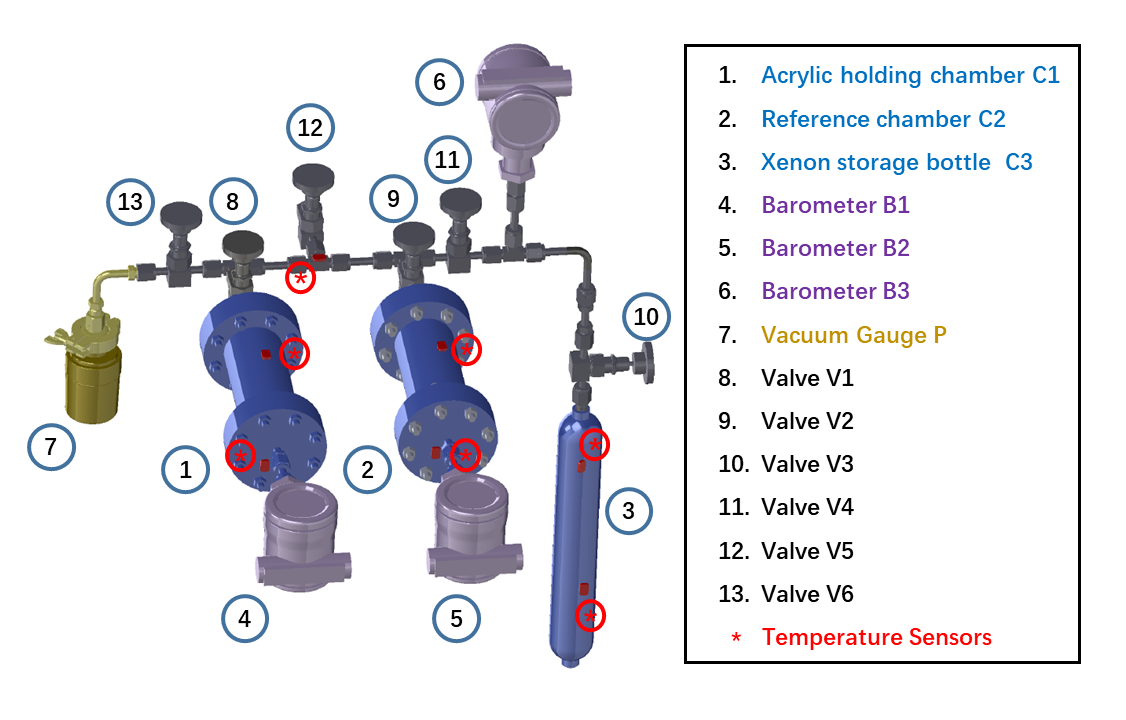}
    \caption{
   The sketch of the system studying the acrylic absorption.
   Marked by red stars are the thermal probes attached to the system surface.
    }
    \label{fig:SystemSchetch}
  \end{center}
\end{figure}

\subsection{Experimental setup}
\begin{figure}[tb]
  \begin{center}
    \includegraphics[width=0.7\linewidth]{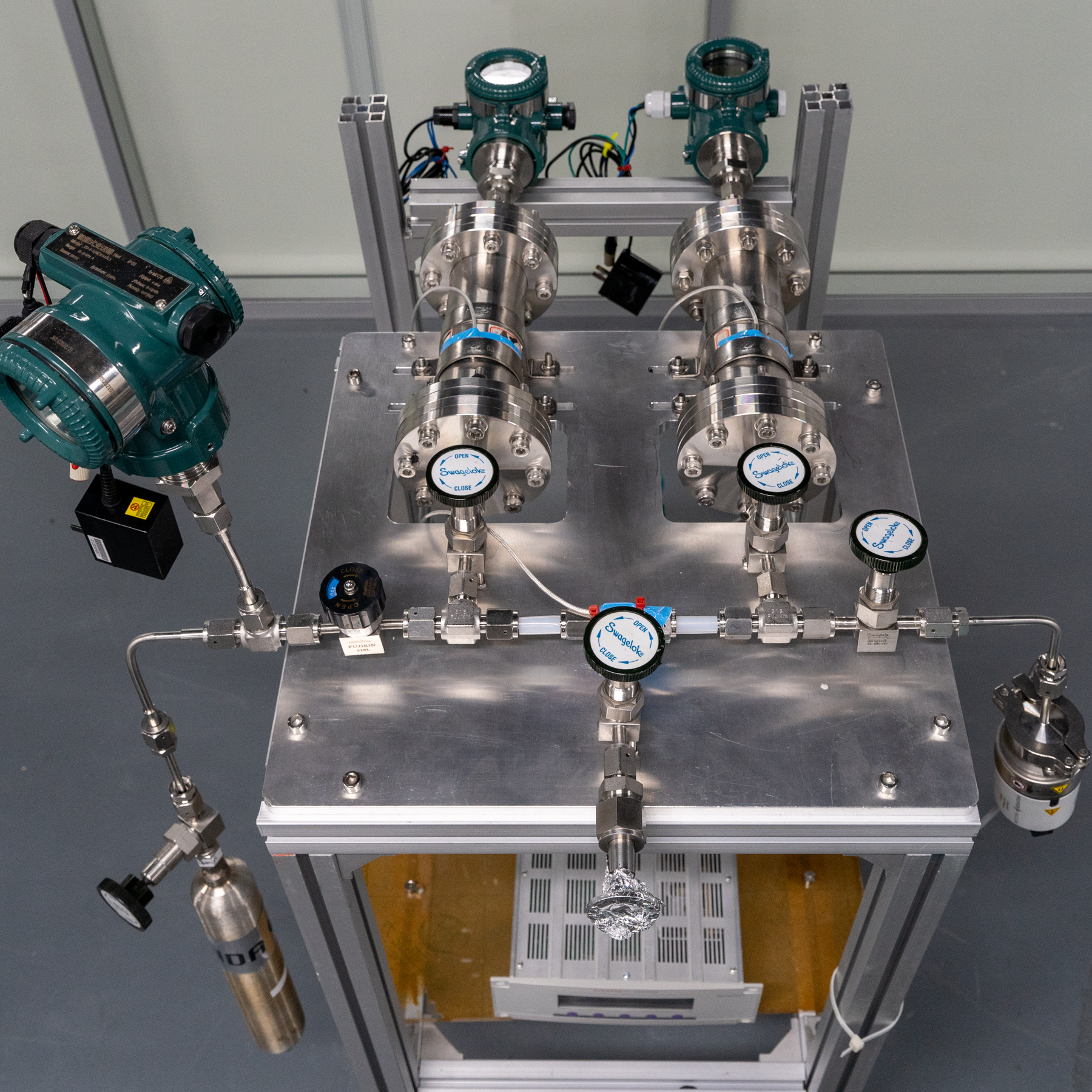}
    \caption{
   The overview of the experimental setup.
   During operation, the setup would be covered with thermal insulation materials to prevent dramatic temperature change.
    }
    \label{fig:SystemSetup}
  \end{center}
\end{figure}

The schematic design and a picture of the setup are shown in Fig.~\ref{fig:SystemSchetch} and Fig.~\ref{fig:SystemSetup} respectively.
Two identical stainless steel (SS) chambers (C1 and C2) are mounted side-by-side.
C1 is used for the absorption measurement where acrylic samples are loaded while C2 works as a reference.
A 0.5~L high-pressure gas cylinder from Swagelok functions as the xenon storage and recuperation chamber.
Three Barometers (denoted as B1, B2, and B3) are mounted to the three chambers for pressure measurements.
All pipe fittings and flanges are metal-to-metal to minimize leakage and outgassing.
A full-range vacuum gauge is also mounted for multiple purposes.
It can be used to measure the inner volumes, monitor the internal outgassing rates, and determine the leakage rates of the systems.
Seven PT1000 temperature sensors are attached to various locations of the system.
The sensors are shown as red stars in the schematic.
The sensors mounted on C1, C2, and C3 provide temperature correction to pressure measurement and others are for redundancy and cross-check.

The system is divided into three segments.
We re-use C1 and C2 to denote the volumes of the measurement chamber and reference chamber.
Each volume consists of the contribution from the vessel, barometer, valve, and their joints.
The two volumes are separated with Valve V1 and V2 from the rest of the inner volume, which is called volume C3.
C3 includes the rest of the pipes, the vacuum gauge, and the storage chamber.
The three volumes can be determined from mechanical designs but more accurately from relative pressure measurement as will be shown later.

\begin{figure}[tb]
  \begin{center}
    \includegraphics[width=0.7\linewidth]{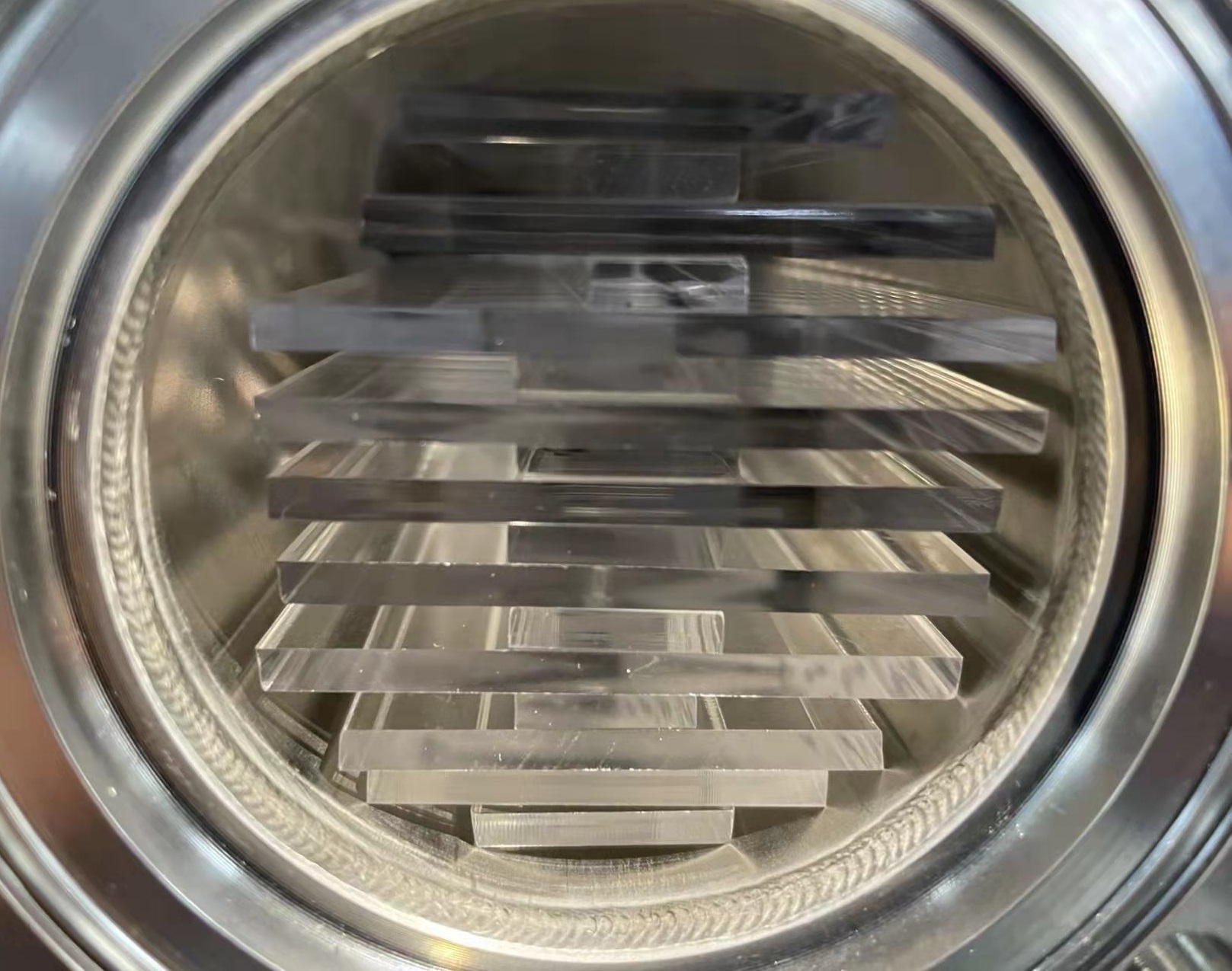}
    \caption{
     The layout of the acrylic samples in C1. The acrylic samples used in the experiment were manufactured by Taixing Tangchen Acrylic Co., Ltd..
    }
    \label{fig:Layout}
  \end{center}
\end{figure}

The accuracies of the instruments used in our measurement are listed in Table~\ref{instrument_accuracy}.
 \begin{table}[htb]
\centering
\caption{Accuracy of the instrument in primal measurements}
\label{instrument_accuracy}
\begin{tabular*}{\columnwidth}{@{\extracolsep{\fill}}llll@{}}
\hline
Instruments & Measurement Range & Accuracy\\
\hline
Barometer B1           & 0 to 3.5~Mpa & 0.075\% of full range\\
Barometer B2, B3        & 0 to 1.6~Mpa & 0.075\% of full range\\
Thermometers           & 0 to 50 $^{\circ}$C & 0.2\%\\
ADC for Barometers   &  & 0.05\%\\
ADC for Thermometers  & & 0.1\%\\
\hline
\end{tabular*}
\end{table}

%

\subsection{Experimental procedures}
\label{sec:PressureEvolution}

\begin{figure}[htb]
  \begin{center}
    \includegraphics[width=1\linewidth]{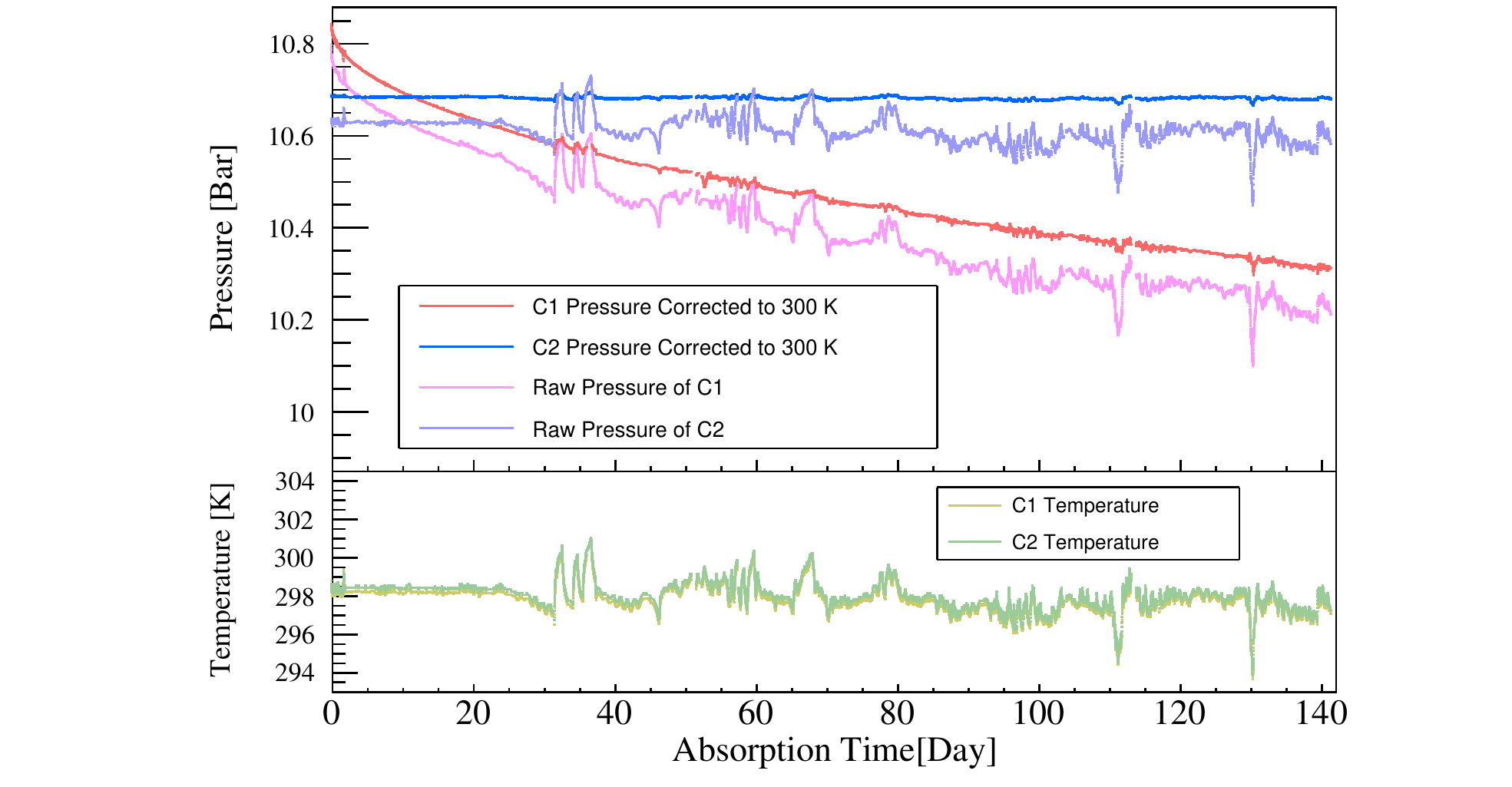}
    \caption{
   (Top panel): the pressure evolution (corrected to values at 300~K) of the vessels with (red) and without (blue) acrylic loaded respectively.
   The raw pressures without correction are plotted in light blue and light red correspondingly.
   (Bottom panel): the synchronous temperature evolution of the two vessels.
    }
    \label{fig:PressureCurve}
  \end{center}
\end{figure}

The major effort of this experiment focuses on measuring the pressure decrease of xenon in the acrylic holding chamber C1 and comparing with the reference chamber C2.
Acrylic sample plates (0.3~cm-thick) were stacked together and inserted into C1, as shown in Fig.~\ref{fig:Layout}.
Meanwhile, the chamber C2 was kept empty as a reference.
The whole system except the storage chamber was pumped to reach a vacuum of 10~mPa at first.
Then chambers C1 and C2 are gradually filled with xenon from the storage chamber.
The filling process lasted less than two minute.
The evolution of pressures of the two identical chambers during one of our measurements is shown in Fig.~\ref{fig:PressureCurve}.
From the figure, it is evident that the pressure of the test chamber kept decreasing, while the reference chamber pressure remained relatively stable.
The pressure drop of C1 seemed to slow down over time but did not show signs of stopping after over 4 months of measurement.
We stopped the measurement on day 142.

Temperature variations were monitored continuously and used to correct the pressure measurement.
The setup was installed in an air-conditioned room and covered with thermal insulation materials to limit the temperature variation.
The short-term fluctuation of temperature was within $1\,^{\circ}$C for most of the time, as shown in the bottom panel of Fig.~\ref{fig:PressureCurve}.

Fig.~\ref{fig:PressureCurve} (top panel) shows the pressures of C1 and C2 corrected to values at 300~K.
The ideal gas model is simply used in the correction, considering the minimal temperature fluctuation.
Overall decreases of $0.529\pm0.017$~bar and $0.004\pm0.001$~bar are observed for C1 and C2 respectively.
The decrease of C2 pressure is smaller than the uncertainty of C1 decrease and ignored in the calculations next.

\subsection{Volume measurement}\label{sec:volume}
\begin{figure}[tb]
  \begin{center}
    \includegraphics[width=0.5\linewidth]{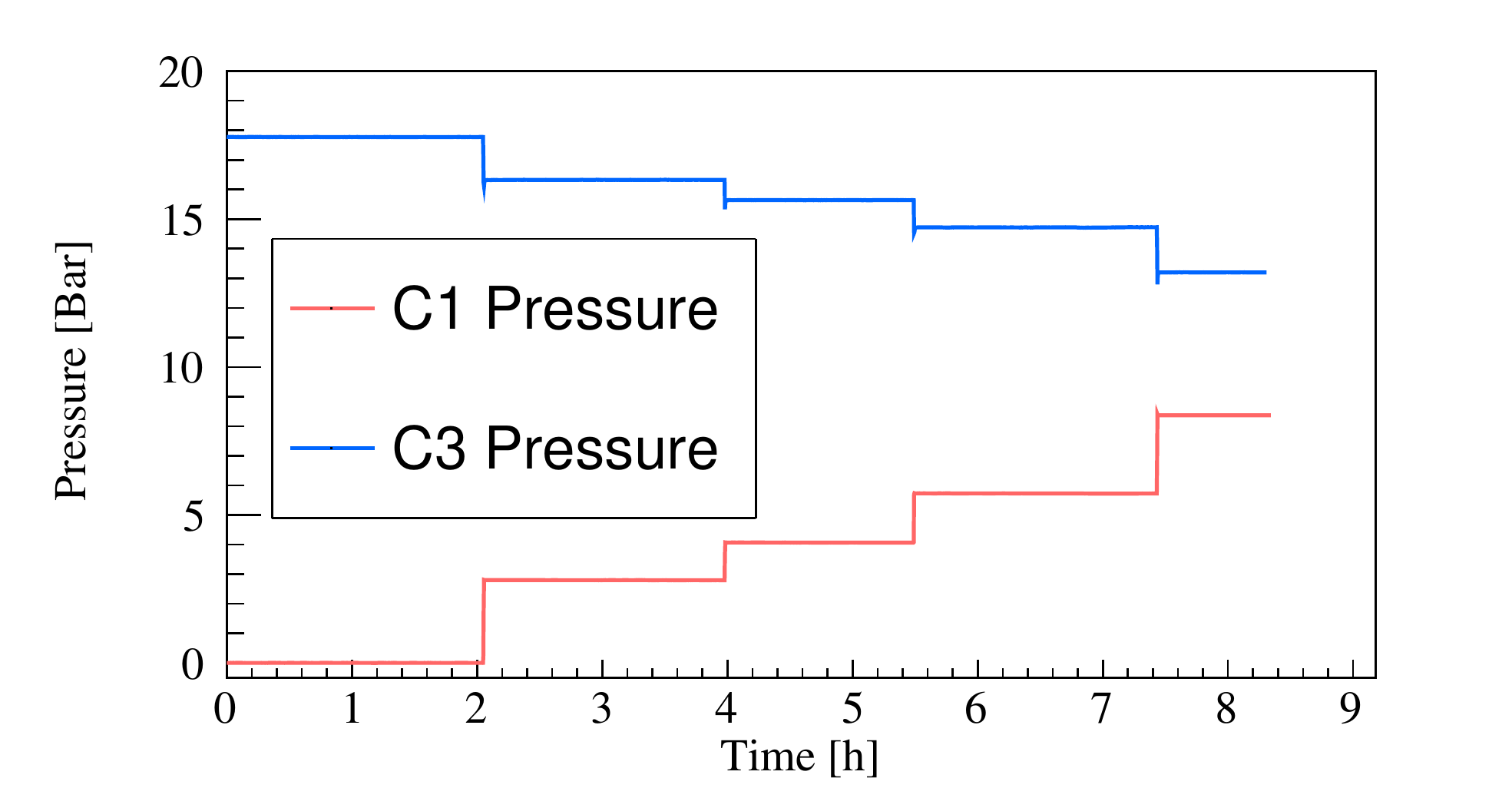}
    \hspace{-5mm}
    \includegraphics[width=0.5\linewidth]{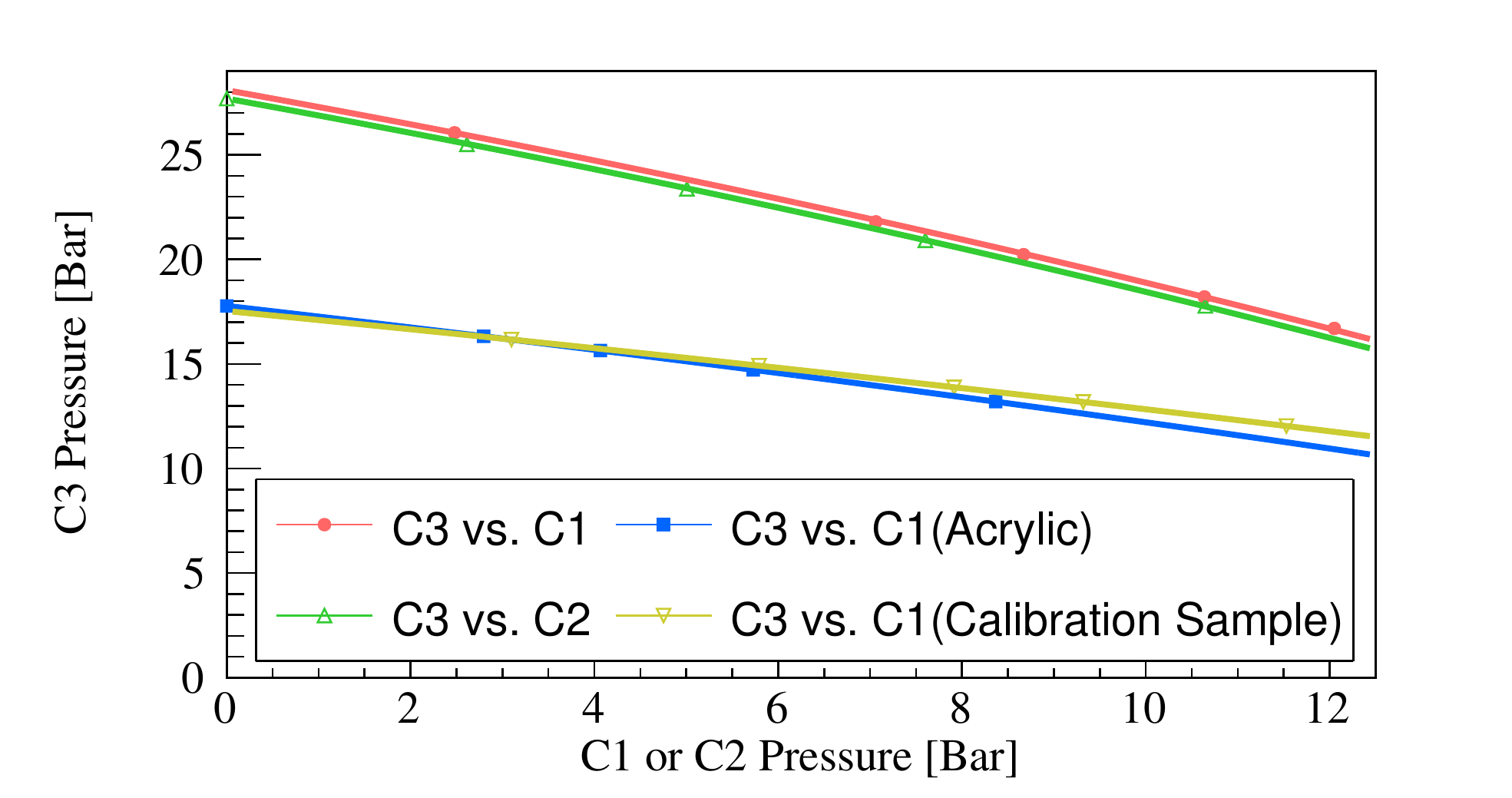}
    \caption{(Left): Example of the pressure changes when gas was transported in the system.
    In the figure xenon gas was filled from sub-spaces C3 to C1 (with acrylic loaded) and the pressure change is plotted as a function of time.
    (Right): Pressures of C3 are drawn as functions of the C2 or C1 pressure.
    The discrete data points (blue) are obtained from the pressure steps of the left figure.
    The corresponding fit gives the result: C1-Acrylic = 0.6319(14)C3.
    Similar fits to determine other volume ratios are also plotted.
    }
    \label{fig:VolumeMeasurement}
  \end{center}
\end{figure}

Precise volumes of the segments were measured to determine the absorption amount in terms of mass, together with the aforementioned pressure decrease.
By transporting gas between segments and monitoring the pressure changes, the ratio of their volumes can be obtained according to the equation of state.
An example is shown in Fig.~\ref{fig:VolumeMeasurement} (left).
We started by filling 17.8~bar of xenon gas in C3 while C1 is under vacuum.
The valve V1 connecting C1 and C3 was opened slightly to allow xenon gas flush to C1.
The flushing was performed with several intermediate steps before the whole system reached equilibrium eventually.
The flushing process, as shown as a step-wise decrease (increase) of pressure in C3 (C1), was carried out slowly to minimize the temperature change due to gas expansion.
The pressure at each step was calculated as the average of measurements in a period with stable temperature.
Based on these values, pressures of C3 are plotted as a function of the C1 pressures in Fig.~\ref{fig:VolumeMeasurement} (right, blue dots) to obtain the ratio of their volumes by fit.
The fit is performed based on the gas equation of state in virial form~\cite{Hurly:1997}:
\begin{equation}\label{EOS}
 P = RT(\rho+B\cdot\rho^2)
\end{equation}
where $P$,R,T,$\rho$ are pressure, molar gas constant, temperature, and density respectively.
B is the second density virial coefficient.
We take the value $B = 0.00103$~[L/g] (at 300~K) from semi-empirical models~\cite{Kaplun:2018} based on experimental data of gaseous xenon in a broad range of pressures~\cite{Michels:1956}.
Such nonlinear term leads to a 6\% deviation on pressure at 10~bar from the idea gas model result, and it is in good agreement with the data.

Similar runs of gas transfer were carried out for different segment pairs.
A solid SS rod (calibration sample) with a volume of 332.5(1)~mL was loaded into the system to determine the absolute magnitudes of all the volumes.
The ratios fitting all the pressure pairs are shown in Fig.~\ref{fig:VolumeMeasurement} (right), and the volume ratios and magnitudes are listed in Table~\ref{VolumeResult}.

\begin{table}[tb]
\centering
\caption{Ratios between volumes and their absolute magnitudes.}
\label{VolumeResult}
\begin{tabular*}{\columnwidth}{@{\extracolsep{\fill}}llll@{}}
\hline
Ratios between volumes & Volume magnitudes (mL)\\
\hline
C1-Acrylic = 0.6319(14)C3 & C1 = 595.9(1.8) \\
C1-Calibration Sample = 0.5349(13)C3 & C2 = 593.3(1.8) \\
C1 = 1.2100(14)C3 & C3 = 492.4(1.4) \\
C2 = 1.2049(15)C3 & C1-Acrylic = 311.2(1.1)mL \\
Calibration Sample = 332.5(1)mL \\
\hline
\end{tabular*}
\end{table}

\subsection{Results and systematics}\label{sec:Outgas}

\begin{table}[tb]
  \centering
  \caption{Systematics of the measurement on gas in C1.}
  \label{ErrorTerm}
  \begin{tabular*}{\columnwidth}{@{\extracolsep{\fill}}llll@{}}
  \hline
  \multicolumn{1}{@{}l}{Systematics Terms} & Quantity[bar$\cdot$mL]\\
  \hline
  SS inner wall outgas  & 3.5 \\
  Acrylic outgas  & 18.9 \\
  Leakage & \textless2.3$\times$10$^{-5}$ \\
  Instrumental accuracy & 4.67\\
  \hline
  \end{tabular*}
  \end{table}

As shown in Fig.~\ref{fig:PressureCurve}, the pressure of the acrylic loading chamber C1 decreased from 10.84~bar to 10.31~bar (corrected to values at 300~K) after 3393.6 hours.
By Eq.~\ref{EOS} the densities of the initial and the final states are calculated: 60.65~g/L and 57.51~g/L.
With the measured gas volume in C1, we can calculate the xenon mass loss as $0.98\pm0.04$~g.
The error shown here only takes into account the instrumental accuracy and statistical uncertainty.
In the following systematic uncertainties are discussed and listed.

The measured pressure change due to xenon absorption in acrylic could be affected by various kinds of gas exchanges, including the outgassing of acrylic samples, outgassing and sorption of inner walls of the system, as well as external and internal leaks.
We calibrated each possible contributions with dedicated runs.
Up to 10~\% of the pressure change was possibly not contributed by the loss of the absorbed xenon.
Table~\ref{ErrorTerm} summarizes all the systematics terms.

\begin{figure}[tb]
  \begin{center}
    \includegraphics[width=0.49\linewidth]{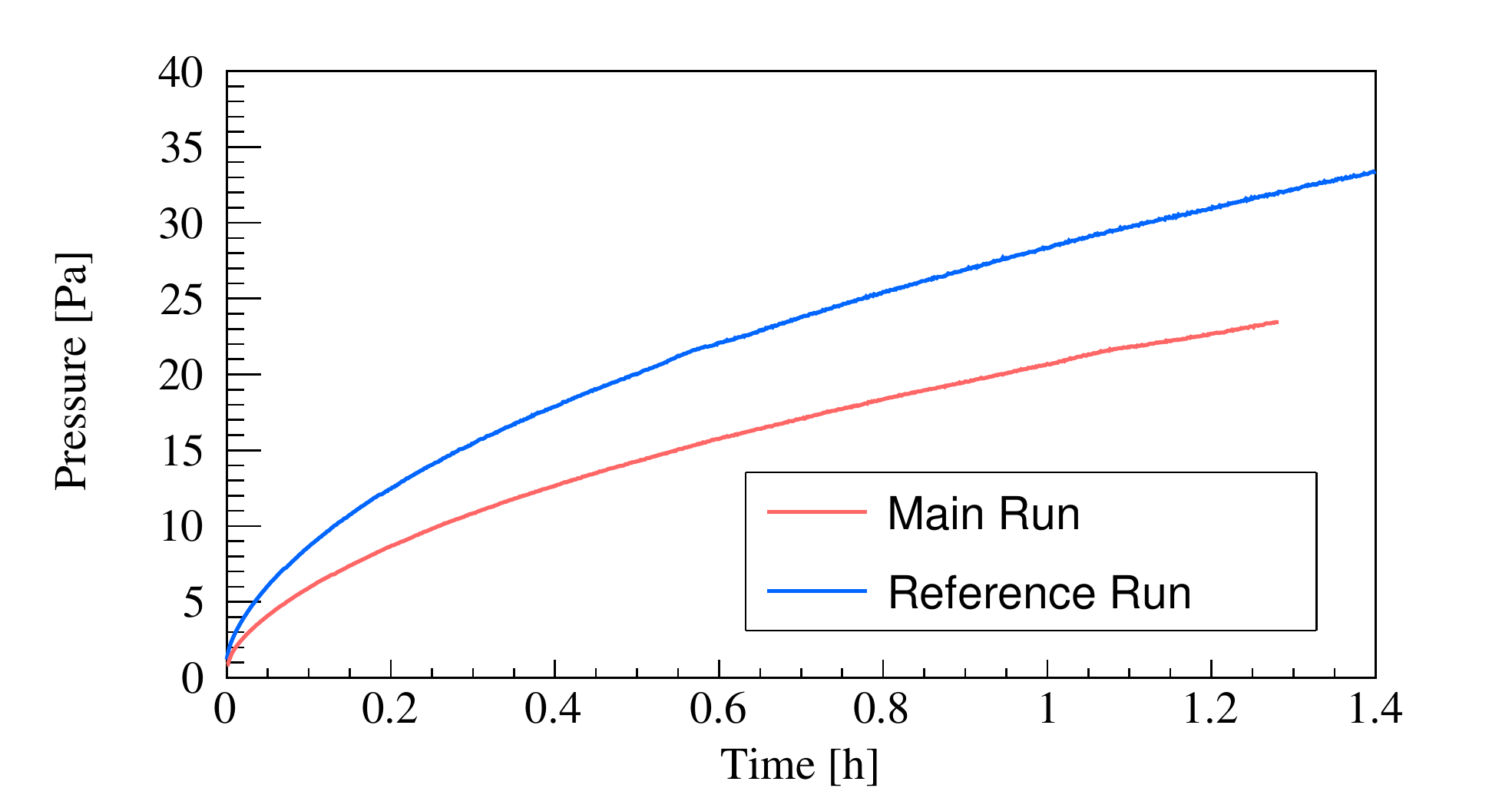}
    \includegraphics[width=0.49\linewidth]{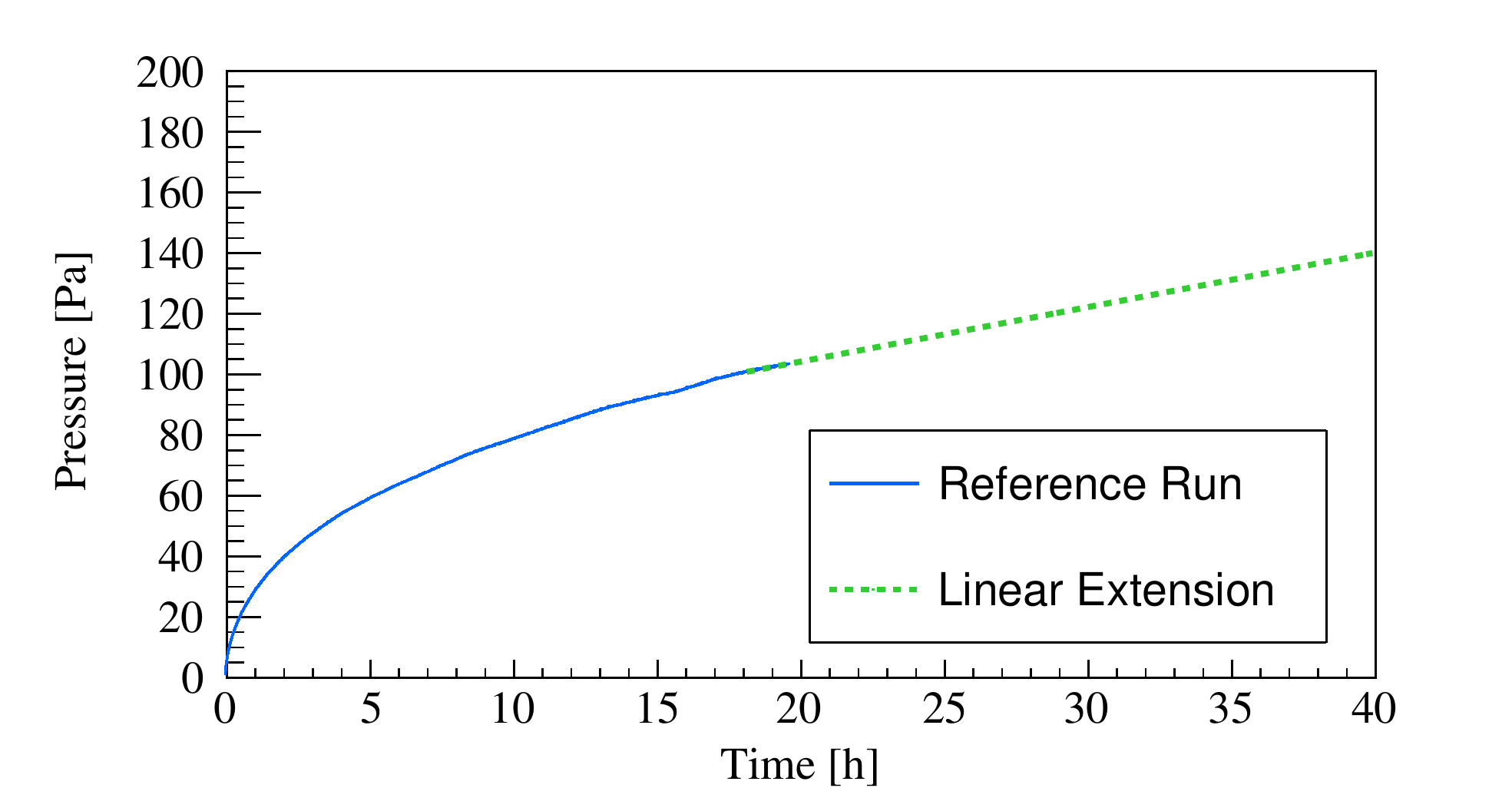}
    \caption{
   (Left): The increasing pressure of C1 characterizes the outgassing effect (dominated by acrylic).
   After the system was pumped to vacuum, the pressure increment was recorded by the vacuum gauge when it was connected with chamber C1.
   During the time V2,3,4,5 were closed and V1,6 were kept open.
   The red outgassing curve is measured 1.5~h before the xenon absorption (main run).
   The longer blue curve (reference run), whose measurement was performed in an earlier time, is used to set a constraint on the outgassing value of the main run.
   (Right): The outgassing curve of the reference run is extrapolated linearly to obtain a conservative approximation of the final outgassing amount.
    }
    \label{fig:AcrylicOutGassing}
  \end{center}
\end{figure}

The most prominent systematics is the acrylic outgassing.
Acrylic outgassing (mostly water vapor) adds an unwanted increase to the measured pressure.
It typically takes two weeks to pump moisture out of acrylic samples till the outgassing rate is sufficiently low.
The outgassing curves right before the absorption measurement (main run) are shown in Fig.~\ref{fig:AcrylicOutGassing}, together with a reference run result when the vacuum level of C1 holding the same acrylic samples were measured for a longer time.
The amount of outgassing in the main run is calculated by linearly extrapolating the pressure curve during the last hours of the reference run.
A conservative estimation of the pressure increase due to the acrylic outgassing is 0.0607~bar (18.9~bar$\cdot$mL with the volume multiplied).
We should emphasize that such calculation is conservative and only indicates the upper limit of outgassing.
With the outgassing curve measured for a longer time, a more precise approximation of the outgassing amount could be made.

The outgassing of SS inner walls of the system can affect the measurement as well, but with minimal contribution compared with the acrylic outgassing.
The outgassing curve, calibrated by the pressure increase of the vacuumized C2 over time, gives an outgassing value of 3.5~bar$\cdot$mL with a similar estimation strategy elaborated above.

Another possible contribution to the systematics is the leakage to the atmosphere from the system when HP xenon is filled, and the leakage between each segment.
To minimize the external and internal leakage in the first place, all the pipe-fitting are metal face-to-face VCR connections, and all flanges are CF type with copper gaskets.
Commercially available valves, Swagelok SS-4BK-V51, are used in the system to separate each segment.
A Pfeiffer GSD320 gas analyzer with a detection limit of 1~ppm was used to ensure no significant external leakage from the chambers filled with 11~bar xenon.
The influence of leakages can be well constraint by the data of C2 pressure, which decreased 0.004~bar after 142~days.
It is assumed that chamber C1, which is instrumentally identical to C2, is also gas-tight and the leakages can be negligible.

In summary, we listed in Tab.~\ref{ErrorTerm} the systematics discussed above, with the instrumental accuracy also shown for reference.
With the systematics added, the amount of absorbed xenon estimated by the pressure drop is 0.98$^{+0.15}_{-0.04}$~g

\subsection{Absorption Dependance on the Sample Geometry}\label{sec:AreaEffect}

\begin{figure}[htb]
  \begin{center}
    \includegraphics[width=0.9\linewidth]{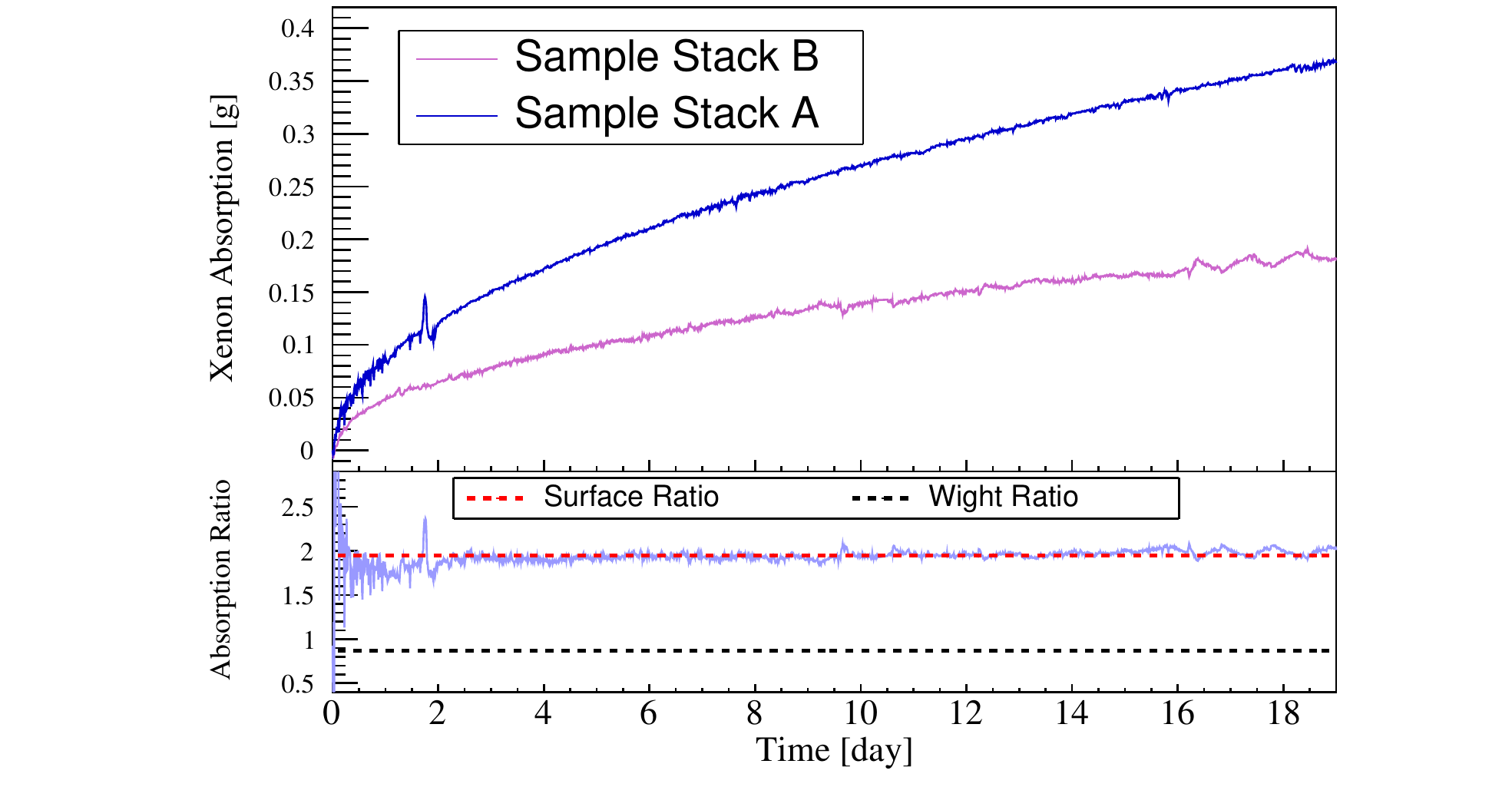}
    \caption{(Top panel): The absorbed xenon in acrylic sample stack A (blue) and B (purple).
    The peak of the blue curve on day 2 is due to abrupt temperature fluctuation caused by accidental air conditioner malfunction, which induces error in the temperature correction.
    (Bottom panel): The ratio of sorption amount by stack A over B.
    Drawn with the dash lines are the corresponding ratios of the weight (black) and surface area (red).
    }
    \label{fig:Compare}
  \end{center}
\end{figure}

To further study the absorption dependence on the sample geometry, another stack of acrylic sample plates (denoted by stack B) with different weight, thickness, and surface area were also measured following the same procedure of the main run stack (denoted by stack A).
This comparison run lasted for 19~days.
The total weight of acrylic stack A (B) was 333.0~g (382.0~g) and the surface area was 2147.2~cm$^2$ (1101.6~cm$^2$).
For each acrylic plate in stack A (B), the nominal thickness was 0.3~cm (1.0~cm).
We intentionally kept the total masses of the two stacks similar while the surface areas were different to compare the effect of the two parameters.
For each stack, the mass of absorbed xenon in acrylic was calculated from the pressure drop and volume.
The amount of absorbed xenon by samples in the two runs are drawn in the top panel of Fig.~\ref{fig:Compare} as a function of the contact time, which is defined as the time relative to the moment we filled xenon gas to the chambers.
For stack A (B), 0.42~g (0.22~g) of xenon was absorbed after 19~days of contact time.
During the time span of our measurement, the pressure of C1 did not approach a stable state, which indicated that the diffusion of xenon into acrylic did not reach equilibrium.

The bottom panel of Fig.~\ref{fig:Compare} shows the mass ratio of absorbed xenon in stack A and stack B.
We added two dashed lines, one for the mass ratio of two stacks and one for the surface area ratio, for visual guidance.
The ratio became approximately constant after 2 days of contact time, and a dominating correlation between the sorption rate and the surface area was observed.

\section{Absorption measurement by weight}
\label{sec:weight}
To cross-check the result based on the pressure measurement, weights of the acrylic samples were measured before and after the xenon absorption measurement.
Firstly the acrylic samples were put in C1 and pumped down.
After the majority of the moisture/air absorbed by acrylic had been pumped out, we measured the initial weight $W_1$ before xenon absorption.
The same samples, together with the C1 chamber, were then fast pumped down again and followed by pressure drop measurement.
At the end of the pressure drop measurement, we took out the acrylic samples to measure the weight $W_2$.
The weight difference $W_2-W_1$ is expected to be mostly from the absorbed xenon gas.

All weight measurements are corrected for air absorption when exposed to air during the measurement period.
Generally speaking, more air could be absorbed by a sample with a larger surface area and in an environment with higher humidity.
However, it is challenging to quantitatively evaluate the correlation.
Therefore, the correction was done in situ.
The samples were grouped into six batches for cross-comparison among the measurements.
Two analytical balances were used to weigh all the 6 groups and each group was measured three times on each balance during a time period of fewer than 30 minutes.
With three measurements done on each batch of acrylic samples at different times, we can compare the weight differences and extrapolate the weights right out of the vacuumized/pressurized chamber.
The summed correction for 6 batches is 0.0138~g (0.0059~g) for the first (second) balance.
We take the variation of the correction of the 6 batches as an uncertainty.

The residual amount of moisture/air in the acrylic sample contributed to the measurement uncertainties.
It is impossible to quantify the residual gaseous impurity when measuring $W_1$ and $W_2$.
Therefore, we aimed to have consistently the same amount of impurity as indicated by the outgassing curves of acrylic samples in the vacuumized C1.
We performed two extra weight measurements after the samples were pumped down in chamber C1.
The outgassing rates were measured and tuned to the same level by adjusting the pumping time.
The two weight measurements give a small discrepancy of 0.0009~g, much smaller than other uncertainties.

When measuring $W_2$, xenon absorbed in acrylic may emanate to the atmosphere and $W_2$ measurement was underestimated.
The emanation could not be measured directly but we could give a conservative estimation.
Ignoring the humidity difference, moisture/air absorption rates calibrated in the $W_1$ measurement (4.5$\times10^{-7}$~g/min$\cdot$cm$^2$) and in the $W_2$ measurement (1.7$\times10^{-7}$~g/min$\cdot$cm$^2$) should be the same and the their difference can be attributed to xenon emanation.
This strategy estimates a total xenon emanation amount of 0.0130~g that we quote as a systematic uncertainty of our measurement.
It should be pointed out that this is an over-estimation since the ambient relative humidity values were 48.2\% and 31.5\% when measuring $W_1$ and $W_2$ respectively, which also contribute to the difference.

The final moisture/air absorption corrected $W_1$ and $W_2$ values for each batch measured by the first balance are listed in Table~\ref{weight}.
The final mass increment results from the two balances are 0.9042$^{+0.0174}_{-0.0044}$~g and the other of 0.8875$^{+0.0117}_{-0.0026}$~g.
The instrumental accuracy, statistical uncertainties, and the systematics discussed above are included in the error.

There is tension among the pressure-drop method and weight measurements due to possible unaccounted systematics.
Further investigation will be needed.

\begin{table}[tb]
\centering
\caption{Weights of the sample groups before and after the xenon absorption experiment. The weight increment ratios by surface area are listed.}
\label{weight}
\begin{tabular}{ccccc}
\hline
\multirow{2}{*}{\shortstack{sample\\ group}} & \multirow{2}{*}{\shortstack{weight \\ before absorption (g)}}& \multirow{2}{*}{\shortstack{weight \\ after absorption (g)}}& \multirow{2}{*}{increment (g)}& \multirow{2}{*}{\shortstack{increment ratio \\ ($\times10^{-4}$~g/cm$^2$)}}\\
&&&&\\
\hline
1 & 36.9161 & 37.0262&0.1101&4.4\\
2 & 52.9983 & 53.1493&0.1510&4.1\\
3 & 84.6154 & 84.8330&0.2176&4.0\\
4 & 28.8878 & 28.9647&0.0770&3.8\\
5 & 51.5625 & 51.6897&0.1272&4.0\\
6 & 77.9406 & 78.1620&0.2213&4.8\\
sum & 332.9207 & 333.8249&0.9042&4.2\\

\hline
\end{tabular}
\end{table}


\section{Result and Discussion}
\label{sec:Conclusion}

The measurement of acrylic absorption in 11~bar xenon was carried out with the xenon pressure drop method and the acrylic weight increment measurements.
Both measurements show clear evidence of xenon absorbed in acrylic samples.
The loss of the xenon pressure indicates an absorption amount of 0.98$^{+0.15}_{-0.04}$~g and correspondingly increments of 0.90$^{+0.02}_{-0.004}$~g (first balance) and 0.89$^{+0.01}_{-0.003}$~g (second balance) are observed on the acrylic weight, for a total of 333 g acrylic samples with a surface area of 2147.2~cm$^2$.
Within the time scale of 142 days of xenon-absorption, the process showed no clear trend of saturation.
The xenon was absorbed by the acrylic samples at a rate of 0.01~g/day at the end of the observation.

When being absorbed into the acrylic, gas particles slowly diffuse in the polymer until the equilibrium state is reached, where the xenon particle density should be uniform in the material.
Therefore the saturated gas absorption amount in acrylic should be proportional to its mass~\cite{Guo:2015}.
Before the saturation, the sorption rate is both influenced by the diffusion coefficient inside the acrylic and the particle transition rate on the gas-polymer interface~\cite{MathOfDiffusion:1975}.
Correspondingly, the absorption rate could be sensitive to either the material thickness or the surface area.
By comparison between different samples, our current measurements show clear dependence of the absorption rate on the surface area in the first 19~days, indicating that the xenon particle transition process on the interface is slower than the diffusion process inside acrylic.
It is expected that such dependence could be reasonably extended since saturation was not observed within the time span of the main run measurement (142 days).

An Experiment in the longer term is to be carried out to study the saturated absorption amount of xenon in acrylic.
Acrylic samples with different geometrical features will also be tested, which could help understand the physics in the absorption process.

The absorption rate of xenon into acrylic normalized to mass or surface area is slower than other gases (i.e. CO$_2$ reported in~\cite{Guo:2015}).
The rate also varies with the target polymer.
Much larger absorption rates of xenon into polymers like polytetrafluoroethylene, high-density polyethylene, etc. were reported in~\cite{NEXT:2018wtg}, and it was indicated that xenon is less active with polyether ether ketone and polyoxymethylene.
Our measurement implies that acrylic behaves relatively stable with xenon despite its well-known porosity and active interaction with many other gases.
It provides positive information for the design of low background detectors, which could contain both xenon and acrylic.
For example, PandaX-III experiment~\cite{Chen:2016qcd} implements 350~kg acrylic with a surface area of 1.2$\times$10$^5$~cm$^2$ as the structural material and 140~kg of xenon at 10~bar as the working gas.
It is estimated that 70~g (scaled by the surface area) of xenon would be absorbed into the acrylic of the detector in the first half-year.

\bibliographystyle{unsrt}
\bibliography{XenonAbsorption}
\end{document}